\documentstyle[aps]{revtex}

\begin{document}

\title{Instability of the symmetric Couette-flow in a granular gas: 
hydrodynamic field profiles and transport}
\draft
\author{M. Sasv\'ari$^{(1,2)}$ J. Kert\'esz$^{(1)}$ and D.E. Wolf$^{(2)}$}
\address{
$^{(1)}$ Department of Theoretical Physics, Technical University of Budapest,
Budafoki \'ut 8, H-1111 Hungary\\
$^{(2)}$ University Duisburg, D-45478 Duisburg, Germany}
\maketitle

\begin{abstract}
We investigate the inelastic hard disk gas sheared by two parallel
bumpy walls (Couette-flow). In our molecular dynamic simulations we
found a sensitivity to the asymmetries of the initial condition of the
particle places and velocities and an asymmetric stationary state,
where the deviation from (anti)symmetric hydrodynamic fields is
stronger as the normal restitution coefficient decreases. For the
better understanding of this sensitivity we carried out a linear
stability analysis of the former kinetic theoretical solution [Jenkins
and Richman: J. Fluid. Mech. {\bf 171} (1986)] and found it to be
unstable. The effect of this asymmetry on the self-diffusion
coefficient is also discussed.

\end{abstract}


\section{Introduction}

In the last decade the clustering instability of undriven granular
gases was extensively researched. Since the first explanation of the
instability \cite{GZ} investigation had been done with several
methods including stability analysis of the hydrodynamic equations
\cite{McN,BMM} and fluctuating hydrodynamics \cite{NE}. The examination of
the influence of nonlinear coupling between hydrodynamic modes was also
performed \cite{BMC}, and the long time
behavior of the clustered state was with mode-coupling theories 
examined \cite{BE}. 
The evolution of vortex velocity patterns preceding the clustering instability 
is well understood \cite{NE}, and the minimal system size, where the
instability appears, and its dependence on the restitution coefficient are
revealed \cite{BMC,J}. 

A driven configuration, the uniformly sheared inelastic hard sphere 
gas (with periodic boundary conditions) shows also an instability and a 
resulting pattern
formation.  In \cite{TG} it was shown that this pattern formation was
guided by an instability for the short times and a shearing caused
convection for longer time-scales. The special importance of this
system is the coupling of macroscopic and microscopic length-scales
in sheared situations as discussed in \cite{SGN,SG,TG2}.

A more realistic configuration is the Couette-flow, the shearing
of the gas between two parallel walls moved in opposite directions.
For this configuration with walls consisting of disks Jenkins and
Richman derived boundary conditions for the momentum and heat transfer
of the walls \cite{JR} proceeding from their kinetic theory for dense
granular gases \cite{JRTh}. In subsequent years this theory was
developed further \cite{RCh} for non Maxwellian velocity
distributions as in \cite{JR}. In all of these papers the resulting
hydrodynamic fields are symmetric (density and granular temperature)
or antisymmetric (flow field) on the half-point between the two
bounding walls. In view of the clustering instability it is of major
interest whether this instability occurs in this configuration and, 
if so, what are its
consequences in the Couette-flow. This paper is devoted to the
investigation of these questions.

The paper is organized as follows:  In the next Section (\ref{simul})
we present molecular dynamic simulations of this system where we found
an interesting sensitivity to the initial condition of the
simulation and asymmetric hydrodynamic fields contradicting the
results in \cite{JR}. Therefore we carried out the linear stability
analysis (Sec. \ref{jrstab}) around the solution of Jenkins and
Richman \cite{JR} and
found it to be unstable against certain fluctuations. We discuss the
effect of this instability on the diffusion coefficient in
Sec. \ref{diffcoeff}  and close the paper with summary in
Sec. \ref{disco}. 

\section{Simulation Results}
\label{simul}
The system considered here consisted of $N$ identical, inelastic hard
disks with mass $m=1$ and radius $r=1$ confined in a
rectangular area of the size $L_x \times L_y$. This two-dimensional 
area is bounded by two parallel
walls from two sides which also define the direction $x$ and are of the 
length $L_x$.
The system is closed through periodic boundary conditions in the $x$ 
direction. The walls
are $L_y$ distance apart ($y$ direction). The origin of the coordinate
system is placed in the middle of the simulated system what sets the 
center of the wall disks to $y=\pm L_y/2$.

The disks interact through the inelastic rough hard sphere potential
meaning instantaneous two particle collisions characterized by given
ratios of the final and initial velocities in the normal and
tangential direction as the normal and tangential restitution
coefficients $e_n=-v_n^f/v_n^i$ and $e_t=v_t^f/v_t^i$, $f$
and $i$ meaning the final and initial velocities.
In case of sliding contacts $e_t$ is replaced by the ratio of
tangential and normal momentum transfer characterized by the friction coefficient
$\mu$.
As a driving force we move one of the walls with constant velocity
$2 U$.

We investigated the system with the event-driven molecular dynamics
method ideal for simulating instantaneous collision
rules \cite{W}. Characterizing  the collisions we used $e_t=-0.3$ and $\mu=0.2$
and varied $e_n$ between $0.6$ and $0.99$.
We consider the system sized $L_x=L_y=40$ and the wall velocity
$2U=10$ if not otherwise mentioned.
The post-collision velocities and angular velocities as a function 
of the pre-collision ones are (for $m=1$, $r=1$),
\begin{eqnarray}
{\mathbf v}_i^\prime&=&{\mathbf v}_i-\frac{1+e_n}{2} \big( ({\mathbf
v}_i-{\mathbf v}_j ) {\mathbf r}_n \big) {\mathbf r}_n + \frac{j_t}{m}
{\mathbf r}_t \\
{\mathbf v}_j^\prime&=&{\mathbf v}_j+\frac{1+e_n}{2} \big( ({\mathbf
v}_i-{\mathbf v}_j) {\mathbf r}_n \big) {\mathbf r}_n + \frac{j_t}{m}
{\mathbf r}_t \\
\mbox{\boldmath$\omega$}_i^\prime&=&\mbox{\boldmath$\omega$}_i -
\frac{j_t}{I_i} ({\mathbf r}_n \times {\mathbf r}_t)\\
\mbox{\boldmath$\omega$}_j^\prime&=&\mbox{\boldmath$\omega$}_i -
\frac{j_t}{I_j} ({\mathbf r}_n \times {\mathbf r}_t)
\end{eqnarray}
${\mathbf r}_n$ being the normal vector of disk surfaces in collision 
${\mathbf r}_n=({\mathbf r}_j-{\mathbf r}_i) /
|{\mathbf r}_j-{\mathbf r}_i|$ and ${\mathbf r}_t$ the tangential
vector, the normal vector rotated with $\pi/2$ in anti-clockwise
direction. $I_i$, $I_j$ are the moments of
inertia of the disks. The parameter $j_t$ depends on the type of the
collision in tangential direction,
\begin{eqnarray}
j_t&=&-\mu \frac{1+e_n}{2} ({\mathbf v}_i-{\mathbf v}_j ) {\mathbf r}_n
\quad \mbox{if} \quad \phi > \phi_0 \quad \mbox{(sliding contact);} \\
j_t&=&\frac{e_t-1}{7} {\mathbf v}_t 
\qquad \qquad \qquad \ \ \mbox{if} \quad \phi \le \phi_0 \quad \mbox{(sticking contact);} 
\end{eqnarray}
where ${\mathbf v}_t$ is the tangential velocity
\begin{equation}
{\mathbf v}_t=({\mathbf v}_i-{\mathbf v}_j ) {\mathbf r}_t+r(
\mbox{\boldmath$\omega$}_j+\mbox{\boldmath$\omega$}_j)
\end{equation} 
and $\phi_0$ separates the sliding and sticking regions in the angle
of incidence $\phi=|{\mathbf v}_t|/
(({\mathbf v}_i-{\mathbf v}_j ) {\mathbf r}_n)$, 
\begin{equation}
\phi_0={7 \over 2} \mu \frac{1+e_n} {1-e_t} .
\end{equation}

To avoid inelastic collapse \cite{coll},
the intrinsic numerical breakdown of the method, we stopped the
simulations when the time interval between two subsequent collisions
became smaller than the precision of the computations as proposed 
in \cite{GZ};
but this stop occurred only for the used smallest restitution
coefficient $e_n=0.6$ and for special initial conditions where most of
the particles did not move, and the well-known chain-like arrangement  
of particles could
evolve after the simulation started. For the experiments below
we used particle numbers $N=100-240$ meaning area fraction 
$\nu\in[0.245,049]$. As test runs for larger systems we used
\{$L_x=L_y=100, N=500, (\bar{\nu}=0.162)$\} and \{$L_x=L_y=100, N=1000,
(\bar{\nu}=0.324)$\} parameter sets resulting in the same phenomenon.

Using several different starting configurations we noticed that there
appears a dense region at one of the walls in the system which is more
and more distinct with decreasing $e_n$ and not recognizable for $e_n$
near $1$, FIG 1.  Using randomly placed particles and an initial
uniform velocity distribution with several mean values in the $x$
direction we observed that the initial value of the mean particle
velocity determines the wall which is chosen for building up a
clustered regime. If we define $v_d$ as
\[
v_d=\left< v \right> - U
\]
we can characterize the final place as if $v_d>0$ the upper $y=L_y/2$
wall is chosen, if  $v_d<0$ the lower $y=-L_y/2$ one, FIG 2.
We used different initial conditions to test this finding \cite{TGF}.
With particles placed in a stripe in the middle
of the system, organized on a lattice with initial velocity $U$,
and using two of them shot against the two wall with
initial velocities which were Galilei-symmetric with the situation of
two walls moved with $\pm U$ we could maximize the time needed to develop the
non-symmetric density field but it appeared also after this initial
condition because numerical
errors provide the needed fluctuations to drive the system in one of
the steady states.

We did simulations for smooth disks
($e_t=1$ and $\mu=0$) as well
and found the same effect. Namely, if we chose $e_n$ to be small
enough there is a distinct band of particles at one of the walls and
for $e_n$ close to $1$ this band does not evolve.  The normal
restitution below which we can speak about an asymmetric phase depends
on the density as for higher densities it appears at higher $e_n$
values on the same system size $L_y$. It also depends on $L_y$ for
fixed density as for larger $L_y$ the kinetic energy influx per
particle decreases as it is proportionate with the length of the wall.
The assumed critical restitution coefficient $e_c(\rho,L_y)$ does not 
depend on the driving velocity as $U$ is the
only parameter which includes dimensionality of time and changing $U$
is a rescaling of the time unit. For the investigation
of intermediate $e_n$ values for which the dense stripe is not so
definite we measured density and velocity profiles and the granular
temperature in the system and found the same effect now in the
asymmetric nature of this functions FIG 3. One can also observe that
with decreasing restitution the minimum of the granular temperature
moves to one of the walls. 

We found these results valid for
a broad range of densities and for different system sizes. For low
restitution there appeared a crystalline structure as compact layers
near a wall. The fact that the simulations did not stop at such 
dense configuration implies that the disks moved around a fixed place
and did not form a long enough connected line of particles in which inelastic
collapse could occur \cite{coll}.

Simulations with time step driven molecular dynamics using damped
harmonic oscillator forces between the particles as a function of
their overlap \cite{W} -- for achieving velocity independent
restitution -- show the same clustering effect at the walls. This
suggests that this effect is closely related to the dissipative
collisions and does not require instantaneous collision rules.   
 
\section{Stability of the Jenkins-Richman solution}
\label{jrstab}
Our above findings contradict the kinetic theoretical calculation of
Jenkins and Richman \cite{JR} and the improved versions of the problem
\cite{RCh} inasmuch as the cited results are always (anti)symmetrical profiles
in the $y$ direction for the hydrodynamic fields of density, flow
velocity and granular temperature.  This contradiction raises the
question of the stability of the (anti)symmetric solution.
Moreover, it is plausible that in an elastic system the hydrodynamic
profiles are symmetric, though this is not a steady state because of
the ever rising temperature caused by viscous heating. Therefore it is
of interest to see weather a phase transition occurs in the system,
specifically, weather there exists a critical $e_n$ value or the
(anti)symmetric solution becomes unstable at arbitrarily small inelasticity.
Because at high
densities the instability in question arises at normal restitution
closer to $1$ and kinetic theoretical calculations are more and more
inadequate for increasing inelasticity it is reasonable to use
kinetic theories for dense systems and we desist from the
consideration of more sophisticated
hydrodynamic equations for low density systems \cite{BDKS,GD}.
Another reason for taking theories for dense gases into consideration is
that the underlying assumptions are more valid as the ratio of the 
simulated system size to the mean free path is much larger in a dense gas. 
For this reasons we
performed a linear stability analysis of the Jenkins-Richman
solution. We now first recite the equations of Jenkins and Richman
\cite{JRTh} and the 
boundary conditions and the solution of the problem \cite{JR}. Then we
perform a linearization around this solution and calculate numerically
the eigenvalues of the stability matrix.

\subsection{Jenkins-Richman Solution} In this section we briefly 
describe the Jenkins and
Richman \cite{JR} solution for a two-dimensional system of inelastic
hard disks driven by
two parallel bumpy walls.
The hydrodynamic equations for the density $\rho$, flow
velocity ${\mathbf u}$ and
fluctuation energy $T$ for smooth disks without external forces are
according to \cite{JRTh}:
\begin{eqnarray}
\dot\rho &=& - \rho {\bf \nabla \cdot u} \ , \\
\rho \dot {\bf u} &=& - {\bf \nabla \cdot P} \ ,\\
\rho \dot T &=& - {\bf \nabla \cdot Q} - \mbox{\rm Tr}
({\bf P \cdot \nabla u}) -\gamma \ . 
\end{eqnarray}
Here the dissipation rate $\gamma$ and heat transport coefficient
$\kappa$ are of the form, 
\begin{eqnarray}
\gamma={4(1-e_n)\kappa T \over {\sigma^2}} \\
\kappa = {2\rho\sigma\nu g_0 T^{1\over 2} \over \pi^{1\over 2}}
\end{eqnarray}
where $g_0$ is the Enskog correction term accounting for excluded
volume effects as a function of the packing fraction $\nu$:
\[
g_0 = {16-7\nu \over 16 (1-v)^2},
\]
and $\sigma=2r$ is the disk diameter.
The constitutive relations for the energy flux $\mathbf Q$ and
pressure tensor $\mathbf P$ are,
\begin{eqnarray}
{\mathbf Q}&=& -\kappa {\mathbf \nabla} T \\
{\mathbf P}&=&(2\rho\nu g_0 T - {1\over 2} \kappa \mbox{\rm Tr}
({\mathbf D})
{\mathbf I} -
\kappa {\mathbf D} .
\end{eqnarray}
From the boundary geometry and the assumption of a Maxwellian velocity
distribution the momentum and fluctuation energy supplied by the
wall in unit time and length are determinable \cite{JR}. This needs an
expansion in $\epsilon=\sigma/L_y$ and assumptions on the dependence
of given ratios of the hydrodynamic fields on $\epsilon$ (see
\cite{JR}), what narrows the validity of the theory. 
The supplied momentum obtained that way is
\begin{equation}
\label{eq1}
M_\alpha={1\over 2} \rho \chi (1+e_n) T \left( n_\alpha+\left({2\over
\pi} \right)^{1\over 2} \frac{v_\alpha}{T^{1\over 2}} \left(
\frac{\theta}{\sin\theta} - \cos \theta \right) + \left({2\over \pi} \right)^{1\over 2} 
\frac{\sigma}{T^{1\over 2}} u_{\gamma,\beta} I_{\alpha\beta\gamma}
\right)
\end{equation}
with
\begin{equation}
I_{\alpha\beta\gamma}=\left({2\over 3}\sin^2\theta-2\right) n_\alpha n_\beta
n_\gamma -{2\over 3}\sin^2\theta \left(n_\alpha t_\beta t_\gamma +
n_\beta t_\alpha t_\gamma + n_\gamma t_\alpha t_\beta\right)
\end{equation}
and the supplied fluctuation energy is
\begin{equation}
\label{eq2}
D=\left({2\over \pi} \right)^{1\over 2} \frac{\rho \chi(1-e_n) T^{3\over 2}}{\sin\theta}
\end{equation}
where ${\mathbf n}$ and ${\mathbf t}$ are the normal and tangential
vectors of the wall respectively and $\theta$ characterizes the
bumpyness of the wall. For wall disks with the same disk diameter
as of the gas particles it equals \cite{JR}:
\[
\sin\theta = {\sigma \over d}
\]
where $d$ is the distance separating the centers of two wall disks.
For the wall geometry of our considered system $d=2\sigma$ and
therefore $\sin\theta=1/2$ and $\theta = \pi/6$. In
(\ref{eq1},\ref{eq2}) $\chi$ accounts for static correlation effects
in collisions at the boundary and $v_\alpha, \alpha=x,y$ is the slip
velocity, the difference between the velocity of the wall and the flow
$\mathbf v=U-u$. From the equality of momentum and energy
transfer of the wall for unit time and unit length the boundary
conditions are:
\begin{eqnarray}
{\mathbf M} = {\mathbf P \cdot n} \ ,\\
{\mathbf M \cdot v} - D = {\mathbf Q \cdot n} \ .
\end{eqnarray}
The solution can be calculated from these equations except one parameter
-- the value of solid fraction at the boundaries (for fixed number of
particles) -- which must be iterated to get the given density in the system 
what we did numerically. Thus the form of the hydrodynamic fields
solving the boundary value problem is, with $T(y)=w(y)^2$,
\begin{eqnarray}
\label{sol}
w(y)&=& \frac{\lambda(U-v) N \sigma}{2\pi^{1\over 2} \sinh(\lambda/2)
S L} \cosh\left( {\lambda y \over L} \right) \ ,\\
u(y)&=& \frac{U-v}{\sinh(\lambda/2)} \sinh\left( {\lambda y \over L}
\right) \ .
\end{eqnarray}
with $S$ and $N$ being the shear stress and pressure respectively and
$\lambda$ must be solved from \cite{JR},
\begin{equation}
{\lambda \over 2} \tanh\left({\lambda \over 2}\right) =
\frac{(1-e_n)L\theta}{2\sqrt{2} \sigma \sin\theta} \left(
\frac{\sin\theta[1-(4\sqrt{2\sigma}/3\sigma) \sin^2\theta]}
{\theta(\theta/\sin\theta-\cos\theta)} -1 \right).
\end{equation}
The density, calculated from 
\begin{equation}
\chi = 2 \nu g_0
\end{equation}
must fulfill the condition that its integral on the whole system
gives the prepared particle number. As mentioned, in that condition 
the only free parameter remains the density at the boundaries what 
must be calculated iteratively.

\subsection{Stability Analysis}
Around the above solution of the boundary value problem we performed a
linear stability analysis with the perturbed quantities,
\begin{eqnarray}
\rho^\prime &=& \rho + \delta\rho \\
{\mathbf u}^\prime &=& {\mathbf u} + \delta{\mathbf u} \\
T^\prime &=& T + \delta T.
\end{eqnarray}
We split the velocity perturbation in the $x$ and $y$ direction and
use plane waves for the perturbations,
\begin{eqnarray}
\delta\rho &=& \delta\rho_{\mathbf k} exp(-i {\mathbf k \cdot r}) \\
\left( {\delta u \atop \delta w} \right) &=& \left( {\delta u_{\mathbf k}
\atop \delta w_{\mathbf k} } \right) 
exp(-i {\mathbf k \cdot r}) \\
\delta T&=& \delta T_{\mathbf k} exp(-i {\mathbf k \cdot r}).
\end{eqnarray}
With this choice the resulting equations have the form
\begin{equation} \label{stab}
\left( 
\begin{array}{c} 
  \partial_t \delta\rho_{\mathbf k} \\ \partial_t \delta u_{\mathbf k} \\ 
\partial_t \delta w_{\mathbf k} \\
  \partial_t\delta T_{\mathbf k} 
\end{array}
 \right) = -{\mathbf A}
\left( 
\begin{array}{c} 
  \delta\rho_{\mathbf k} \\ \delta u_{\mathbf k} \\ \delta w_{\mathbf k} \\ 
\delta T_{\mathbf k}
\end{array}
 \right)
\end{equation}
where ${\mathbf A}$ is the stability matrix with the elements
given in Appendix \ref{matrix}.

With the help of MAPLE V Release 5 we analyzed the eigenvalues 
of the stability
matrix $\mathbf A$ for different densities, restitution
coefficients, wave numbers and sites. 
On FIG 4-7 we present results for some of the parameter sets which
are representative for the different situations. 
With the choice of the negative sign in
(\ref{stab}) 
a negative eigenvalue means an unstable solution.
All of the shown figures feature at least one eigenvalue 
which is negative in the small wave
number region. This signalizes the unstable fluctuation. The upper zero point
of this eigenvalue nears to the zero wave number as we increase the
restitution coefficient and becomes smaller than $2\pi/L$, the smallest
possible wave number in the system, at a certain value of $e_n$
what depends on the other parameters. 
However, this value is very close to $1$ what suggests that the
symmetrical solution is always unstable linearly and nonlinear effects
can move this transition point to higher inelasticities. 
We analyzed the most unstable direction of the solution also as we plotted
the unstable eigenvalue for a given amplitude of the wave number and
for the whole angle measured from the direction of the mean flow in
anti-clockwise direction as seen in FIG 8,9.
According to the figures the most unstable directions are close to the
$\pm \pi/4$ angles which are the same angles found to be the
governing unstable directions in linear order for the pattern
formation by Tan and
Goldhirsch for the uniformly sheared granular gas \cite{TG}.

\section{Self-diffusion coefficient and clustered flows}
\label{diffcoeff}
In this section we investigate the crossover to asymmetric stationary
states of the system and try to show the effect of the instability 
discussed above on the transport coefficients.
For this reason we measured the mean square displacement of the
particles,
\begin{equation}
\label{msqd}
r_x^2=\left\langle (x(t)-\langle v_x \rangle t - x(0))^2\right\rangle,
 \qquad r_y^2=\left\langle (y(t)-y(0))^2
\right\rangle
\end{equation}
and velocity correlation functions,
\begin{equation}
\label{velcor}
c_x=\left\langle (v_x(t)-\langle v_x\rangle) (v_x(0)-\langle v_x\rangle)\right\rangle, \qquad c_y=\left\langle v_y(t)
v_y(0)\right\rangle
\end{equation}
in the system for stream-wise and perpendicular directions. 
In (\ref{msqd}) and (\ref{velcor}) the brackets $\langle
\rangle$ mean space and ensemble averages over the whole
system. $\langle v_x \rangle$ was calculated from the measured average
displacement of the particles averaged over several (50-100) runs and
over particles. Ensemble averages were carried out the following way. 
For every parameter set we let relax the system in its steady state.
Then measured granular temperature averaged over stripes of width
$1=r/2$ generating the $T(y)$ function. Before every measurement we
perturbed the velocities of every particle with a uniformly
distributed velocity the support of the distribution being
$2\sqrt{T(y)}$, where $y$ was the coordinate of the particle.
After relaxation we carried out the measurement.
The $\langle v_x \rangle$ average velocity was evaluated from a linear
fit on the averaged $x$ displacement of the particles being the
steepness of $x$ as a function of time.

From the mean square displacement for long times we could evaluate the
diffusion coefficient of the particles. We fitted a power law with two 
fit parameters $D$ and $b$,
\[
f(t)=2 D t^b
\]
on the mean square displacement restricted to long times.
If the value of $b$ was up to errors ($\pm 0.02$) near $1$ the system 
was assumed to be in a
diffusive state. After that we took $b=1$ and fitted $f(t)$ again for
$D$. The time the system evolves into this diffusive state
depends on $e_n$ as suggested in \cite{BRCG} for the homogeneous cooling
state, and also depends on the steady state configuration which is
also $e_n$ dependent. This has several reasons. First, to achieve 
this diffusive behavior in $x$ direction
the particles have to experience the characteristics of the system 
-- finiteness and inhomogeneity of the hydrodynamic profiles -- 
therefore have to travel between the to walls several times. 
As $e_n$ decreases and a dense cluster evolves at one wall this
recognition takes more and more time. 
Also the form of the velocity correlation function shows increasing
correlations with decreasing $e_n$ as expected. After a short
($t\approx 10$) fast decay $c_x$ shows an exponential decay up to
times $t\approx 100$ and changes to a slow decay for
later times (small $e_n$) or is relaxed (large $e_n$) FIG 10.
Both the short-time correlations and the characteristic time of the
exponential increases as we decrease $e_n$, along with the earlier 
changeover to the slow decay if it is apparent.

The plot of the mean square displacement versus time
diagram (FIG 11) shows a plateau value of the diffusion coefficient
for larger values of $e_n$. Below $e_c=0.82$ $D$ increases fast with 
decreasing $e_n$.
We suggest that this value marks the transition point where the
particles begin to prefer one of the walls and the appearing dilute
regions allow for a greater mobility for the particles.
Making use of the noticed exponential decay of the velocity
autocorrelation $c_x$ we also estimated $D$ as the integral of 
a fitted exponential,
\[
D\approx A \tau \qquad \mbox{\rm if} \qquad c_x \approx A e^{-t/\tau}.
\]
knowing that it underestimates $D$ as $A e^{-t/\tau}$ is smaller than
$c_x$ for very short and for long times, but it gives nearly the same 
result as the long time behavior of $r_x^2$ above the transition point 
and shows a smaller increase below it (FIG 11). This breakup comes 
from the fact that
under the transition point there appears a slow decay in the velocity 
correlation function after the exponential decay used to approximate it.
The short-time exponential decay characterizes only the dilute part
of the system. The slow decay thereafter comes from the averaged
effects of the exigous diffusion in the dense region and from the
influence of 
particles entrapped or emitted by the dense region from or to the
dilute one. In absence of the asymmetric dense state above the 
transition point $e_c$ only the exponential relaxation appears.

The buildup of correlations is also a cause for increasing times 
needed to reach the diffusive state thus increasing simulation time
what restricted us to smaller
system sizes for dense material as it did not allow fast computations.
Therefore the investigation of the size and density dependence of the 
transition point is left for later studies.

The retrievable information from the mean square displacement in $y$
direction is bounded as the system itself namely it goes to a plateau
value according to $L_y$. This finiteness influences velocity
correlations also. $c_y$ shows only short-time fast decaying
correlations (and there appear anti-correlations for small $e_n$.)
With the methods discussed above we do not noticed any mark of the 
transition (only a small sign as presented in \cite{SKW} for time-step
driven simulations).
 
For more definite insights in the behavior of the system we carried out
measurements specific to the configuration of the stationary states.
We divided the system into stripes parallel to the walls and
restricted the averages in (\ref{velcor}) on particles starting
from these stripes,
\begin{equation}
c_x^i=\left\langle \left(v_x(t)-\bar{v}_x^i(t)\right)
\left(v_x(0)-\bar{v}_x^i(0)\right)
\right\rangle_i, \qquad  c_y^i=\left\langle v_y(t)v_y(0) \right\rangle_i
\end{equation}
where $<>_i$ denotes averaging over particles being in stripe $i$ at
$t=0$ and the ensemble average described above. The velocity
$\bar{v}_x^i(t)$ is the average velocity of particles starting from
stripe $i$ at time $t$. This velocity correlation function
characterizes the stripe more and more as the difficulty to leave
the stripe increases which is the case in the dense layer evolving 
at the wall.
The fact that the velocity correlation function $c_x^i$ does not show a
negative minimum at short times (FIG.10) -- which would be a sign of caging
effects -- suggests, that the particles essentially comoving with the
wall move barely in the $x$ direction compared to one another. Instead
because of the sheared situation they are driven by diagonal
collisions which dominate the motion and smooth out $c_x$. 
There is a negative minimum in $c_y^0$ ($0$ indexing the
stripe nearest to the wall at the cluster) but this includes the
effect of the collisions with the wall particles which give a negative
contribution as also seen on the velocity correlations in the stripe
at the opposite wall. This appears also at restitution coefficients well
above the transition.

\section{Discussion}
\label{disco}
In this paper we considered a granular gas sheared by two bumpy walls.
By means of molecular dynamics simulations we discovered that the
system shows a spontaneous symmetry breaking as it evolves to a 
stationary state which is asymmetric in the hydrodynamic fields with 
respect to the centerline of the system. 
The location where this density maximum (temperature
minimum) is shifted to from the symmetry axes is predetermined by the initial velocity distribution
as the system chooses the wall according to this distribution
as discussed in Sec. \ref{simul}.  During the simulations, measuring
the mean $y$ coordinate of the particles, we did not observe any switch
between the possible two states where the asymmetry
was explicit.  We found in a linear stability analysis that the
symmetrical solution proposed by Jenkins and Richman \cite{JR} is
unstable. In analyzing the solution \cite{JR} one finds that the
description of the asymmetric profiles would need higher order
equations in $\epsilon$ as in \cite{JR}. 
We do not consider this analysis as a strict result
quantitatively. According to \cite{G} the equations of Jenkins and
Richman are not suitable for linear stability analysis because they
neglect the time dependence of the mean free path. But the equations
with the corrected constitutive relations \cite{SG} have the same
dependence on the hydrodynamic field and differ in the $e_n$
dependence only. We assume that this correction terms cannot
stabilize the system but can at most increase the $e_n$ value under which the
solution becomes unstable.

We found a  connection to the unstable directions observed in
uniformly sheared granular gases \cite{TG}. That suggests that not
uniformly sheared systems possess the same instability but in the
Couette-case the pattern formation is repressed by the boundary
conditions. However, the geometry of the boundary conditions (walls)
plays a major role in  stabilizing the flow fields leading to a
stationary state.  In the boundary conditions of Jenkins and Richman
\cite{JR} only static correlations are considered and dynamic
correlations, the role of multiple and correlated re- collisions with
the wall, are neglected.  We measured the number of collision
sequences which consist of collisions of a disk with two wall disks
successively because such collisions provide disks leaving the wall in
a steeper angle what can provide greater contribution to the pressure
at lower granular temperature.  In the asymmetric flow regime the
number of such collisions is clearly higher at the preferred wall than
at the other what supports our assumption. This is also implied by the
fact, that in the final stationary state the minimum of the granular
temperature is only at the wall for very high inelasticities or
densities where we could observe closed layers of particles at the
wall chosen. For moderate densities and restitution the temperature
minimum remains in the bulk of the system, and the clustering process
stops at a stable density profile. Such effect was described for a
constrained homogeneous system in \cite{BMC}.  This boundary effects
are one reason for the different results for  the transition point
obtained from the stability analysis or the measurement of the
self-diffusion coefficient and velocity autocorrelation functions
respectively.  The found effect in the dependency of these functions
on the normal restitution coefficient described in
Sec. \ref{diffcoeff} is well described as a consequence of the
transition. However, the system size used in the simulations is too
small to compare the two  results. We suggest that for larger systems
where the Jenkins-Richman theory is more valid we would obtain larger
values for the transition point but the involved measurements leave
this for later studies. As a final note we would like to mention that 
in systems with elastic
collisions with the wall disks the flow with the same parameters
remains symmetric.

The consequences of this paper are hard to verify in experiments
because of the absence of gravity. However microgravity
experiments for the same configuration are under way \cite{LJ}.
We suggest that the two-dimensional case considered here would be 
realizable with disks on an air table as this would exclude gravity 
effects and would not interfere with the behavior of the disks in 
the crucial horizontal direction. Periodic boundary conditions could 
be achieved with two similar stadion-like chains of disks as performed 
in \cite{LJ}.

\section{Acknowledgements}
M.S (J.K) wants to acknowledge the financial help of the DAAD (AvH
Foundation). This work was
partially supported by OTKA T029985 and OTKA T024004.

\appendix
\section{The linear stability matrix}
\label{matrix}
In the following we describe the matrix elements of the linear stability
matrix. The notations include $\rho=\vartheta\nu$ where $\vartheta=\pi / 4$
for $\sigma=1, \ m=1$ and $\varphi=\partial_y (\nu g_0 (\nu))$.
\begin{eqnarray}
A_{11}&=&u i k_x; \\
A_{12}&=&\rho i k_x; \\
A_{13}&=&\rho i k_y + \partial_y \rho; \\
A_{14}&=&0; \\
A_{21} & = & 2T \varphi i k_x+\frac{2\nu g_0 Tik_x}{\rho} -
{\sigma\over\sqrt{\pi}} \partial^2_y u \left(\frac{\nu
g_0\sqrt{T}}{\rho}+\sqrt{T}\varphi \vartheta\right)- \nonumber \\
&& {}-{\sigma\over\sqrt{\pi}} \frac{\partial_yu}{\rho} \left[
\varphi\sqrt{T}+ \nu g_0 \left(\frac{\partial_y T}{2\sqrt{T}}+
\sqrt{T} i k_y \right)+ \partial_y \rho \sqrt{T} \varphi \vartheta
\right]
- \nonumber \\
&& {}-{\sigma\over\sqrt{\pi}} \partial_y u \vartheta \left(\frac{\partial_y
T\varphi}{2\sqrt{T}}+ \sqrt{T}\partial_y \varphi+\sqrt{T}\varphi i k_y
\right); \\
A_{22}&=&uik_x+\frac{\sigma\nu g_0\sqrt{T}}{\sqrt{\pi}}
\left[3k_x^2+k_y^2 -\left(\frac{\partial_y
\rho}{\rho}+\frac{\partial_y
T}{2T}\right)ik_y\right]-{\sigma\over\sqrt{\pi}} \varphi\sqrt{T}ik_y;\\
A_{23}&=&\partial_y u+\frac{\sigma\nu g_0\sqrt{T}}{\sqrt{\pi}}
\left[2 k_x k_y -\left(\frac{\partial_y\rho}{\rho}+
\frac{\partial_yT}{2T}\right)ik_x\right]
-{\sigma\over\sqrt{\pi}} \varphi\sqrt{T}ik_x;\\
A_{24}&=&2\nu g_0 i k_x -\frac{\sigma\nu g_0\partial^2_y u}{2\sqrt{T\pi}}-
-\frac{\sigma\partial^2_y u}{2\sqrt{T\pi}} \left[\varphi+ \nu g_0 
\left(\frac{\partial_y\rho}{\rho}-\frac{\partial_yT}{2T}+i
k_y\right)\right]; \\
A_{31}&=&2\left(\frac{\partial_y\rho T\varphi}{\rho}+ \partial_y T
\varphi + T\partial_y \varphi\right) \vartheta + 2T\varphi\vartheta i
k_y + \frac{2\varphi T}{\rho} \nonumber; \\
&&{}+ 2 \frac{\nu g_0}{\rho} (\partial_y T +
T i k_y) -\frac{\sigma\partial_y u}{\sqrt{\pi}} \left( \frac{\nu g_0
\sqrt{T} i k_x}{\rho} + \sqrt{T}\varphi i k_x \vartheta \right); \\
A_{32}&=&\frac{\sigma\nu g_0\sqrt{T}}{\sqrt{\pi}}
\left[2 k_x k_y -\left(\frac{\partial_y\rho}{\rho}+
\frac{\partial_yT}{2T}\right)ik_x\right]
-{\sigma\over\sqrt{\pi}} \varphi\sqrt{T} i k_x; \\
A_{33}&=&uik_x+\frac{\sigma\nu g_0\sqrt{T}}{\sqrt{\pi}}
\left[k_x^2+3k_y^2 -\left(\frac{\partial_y\rho}{\rho}+\frac{\partial_y
T}{2T}\right) 3 i k_y\right]-{\sigma\over\sqrt{\pi}} \varphi\sqrt{T} 3
i k_y;\\
A_{34}&=&2\nu g_0 \left(\frac{\partial_y\rho}{\rho} + i k_y\right)
+2\varphi-\frac{\sigma\nu g_0 \partial_y u i k_x}{2\sqrt{t\pi}}; \\
A_{41}&=&\left(\frac{8(1-e_n)}{\sigma\sqrt{\pi}} T^{3\over 2}-
\frac{2\sigma\sqrt{T} \partial^2_y T}{\sqrt{\pi}}\right)
\left(\frac{\nu g_0}{\rho}+\varphi\vartheta\right) \nonumber \\ 
&&{}-\frac{2\sigma\partial_y T}{\sqrt{\pi}\rho} \left[\varphi\sqrt{T} +
\nu g_0\left(\frac{\partial_y T}{2\sqrt{T}}+\sqrt{T} i k_y \right) +
\varphi \partial_y\rho\sqrt{T}\vartheta\right] \nonumber \\
&&{}-\frac{2\sigma\partial_y T\vartheta}{\sqrt{\pi}}
\left[\partial_\varphi\sqrt{T}
+ \frac{\varphi\partial_yT}{2\sqrt{T}} + \varphi\sqrt{T} i k_y \right]
- \frac{\sigma\sqrt{T}}{\sqrt{\pi}}(\partial_u)^2 \left(\frac{\nu
g_0}{\rho} + \varphi \vartheta\right); \\
A_{42}&=&2\nu g_0 T i k_x - {2\sigma\over\sqrt{\pi}} \nu g_0 \sqrt{T}
\partial_yu i k_y; \\
A_{43}&=&\partial_yT+2\nu g_0 T i k_y - {2\sigma\over\sqrt{\pi}} \nu
g_0 \sqrt{T} \partial_yu i k_x; \\
A_{44}&=&u i k_x + \frac{12(1-e_n)\nu g_0 \sqrt{T}}{\sigma\sqrt{\pi}}
\nonumber \\ &&{}+ \frac{2\sigma\nu g_0\sqrt{T}}{\sqrt{\pi}} 
\left[k_x^2+k_y^2-
\left(\frac{\partial_y\rho}{\rho}+\frac{\partial_yT}{2T}\right) i k_y
\right] -{2\sigma\over\sqrt{\pi}} \varphi\sqrt{T} i k_y -
\frac{\sigma\nu g_0}{\sqrt{T\pi}} \partial_y^2 T \nonumber \\ &&{}-
\frac{\sigma}{\sqrt{T\pi}} \partial_y T \left[\varphi + \nu g_0
\left(\frac{\partial_y\rho}{\rho}-\frac{\partial_yT}{2T}\right) + \nu
g_0 i k_y \right] - \frac{\sigma\nu g_0}{2\sqrt{T\pi}} (\partial_yu)^2;   
\end{eqnarray}

\centerline{\bf FIGURE CAPTIONS}

\noindent FIGURE 1: Snapshot configurations in the steady state for 
different
restitution coefficients: (1a) $e_n = 0.7$, (1b) $e_n=0.8$, (1c)$e_n=0.9$ 
Lines from centers of particles indicate the direction and magnitude
of its velocity.

\noindent FIGURE 2: Snapshot configurations in the steady state for 
different initial
velocity distributions ($e_n=0.7$). 2a: $v_d < 0$,  2b: $v_d > 0$.

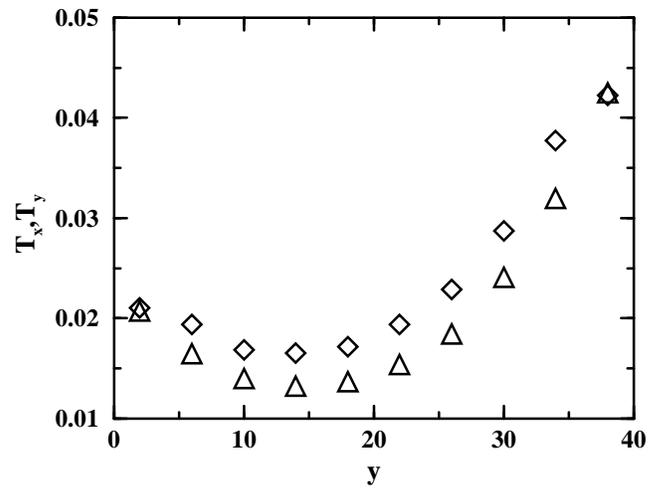
\noindent FIGURE 3: Granular temperature in the $x$ and $y$ 
direction as a function of $y$ (in units of disk radius $r=1$) 
across the system for $e_n=0.9$. Values
are averages over stripes of width 4 parallel to the walls with
diameter $4$ for a system of size $L_y=40$.

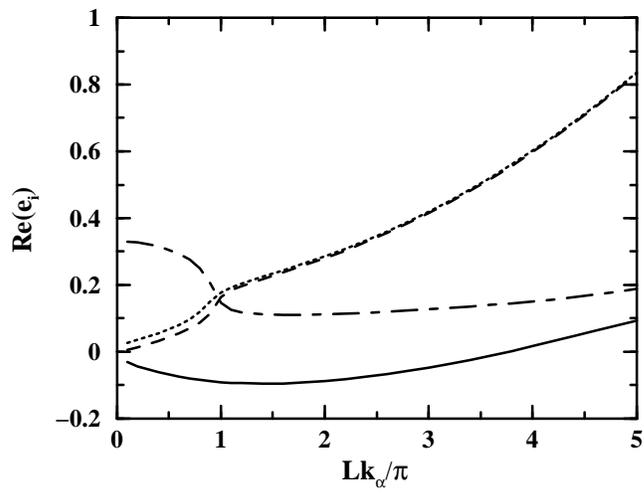
\noindent FIGURE 4: Eigenvalues of the stability matrix for $e_n=0.9$,
$y=0$, average packing fraction $\bar{\nu}=0.486$, and ${\mathbf
k}_\alpha$ points in the direction
$\alpha=\pi/4$ measured anti-clockwise from the mean flow direction. 

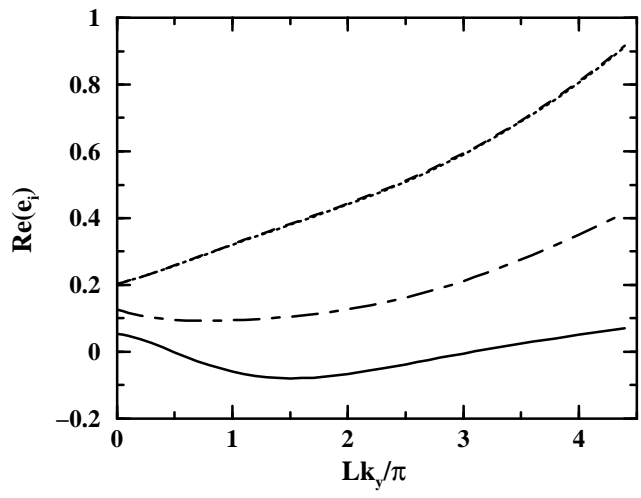
\noindent FIGURE 5: Eigenvalues of the stability matrix for $e_n=0.95$,
$y=0$, $k_x=2\pi/L$ and average packing fraction $\bar{\nu}=0.486$ as a
function of $k_y L/\pi$.

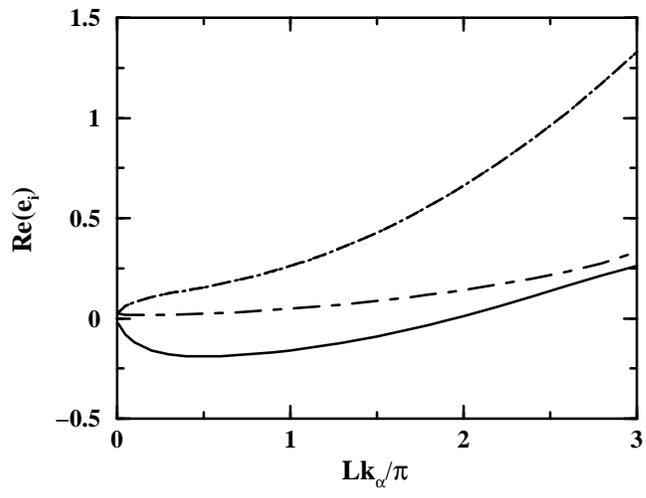
\noindent FIGURE 6: Eigenvalues of the stability matrix for $e_n=0.997$,
$y=0$, average packing fraction $\bar{\nu}=0.486$, and ${\mathbf
k}_\alpha$ points in the direction
$\alpha=\pi/4$ measured anti-clockwise from the mean flow direction. 

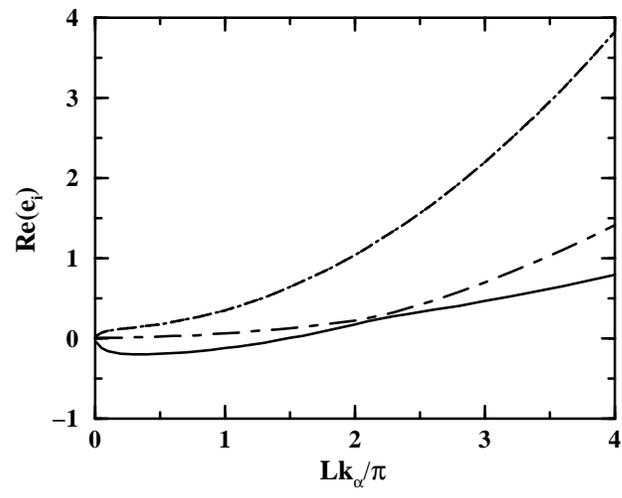
\noindent FIGURE 7: Eigenvalues of the stability matrix for $e_n=0.999$,
$y=0$, average packing fraction $\bar{\nu}=0.486$, and ${\mathbf
k}_\alpha$ points in the direction
$\alpha=\pi/4$ measured anti-clockwise from the mean flow direction. 

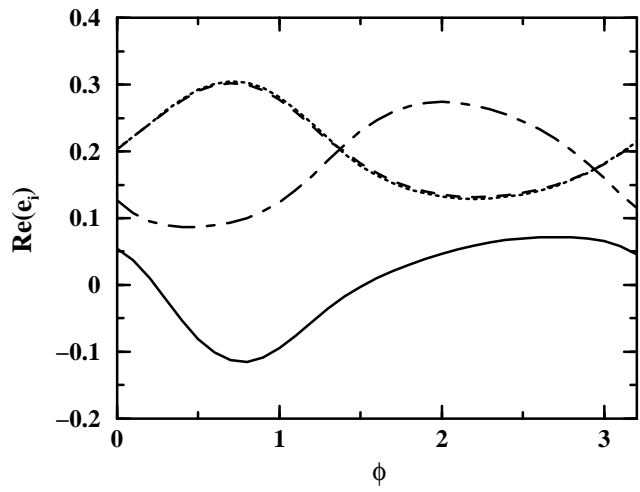
\noindent FIGURE 8: Angle dependence of the real part of the
eigenvalues for
$e_n=0.95$, $y=0$ and $\bar{\nu}=0.486$. The eigenvalue is plotted as a
function of the angle measured from the flow direction 
$k_x=(2\pi/L) \cos(\phi)$ and $k_y=(2\pi/L)\sin(\phi)$.

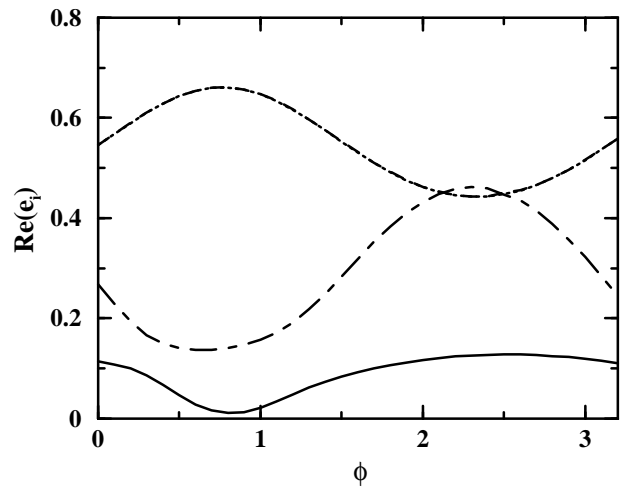
\noindent FIGURE 9: Angle dependence of the real part of the 
eigenvalue for
$e_n=0.997$, $y=0$ and $\bar{\nu}=0.486$. The eigenvalue is plotted as a
function of the angle measured from the flow direction 
$k_x=(2\pi/L) \cos(\phi)$ and $k_y=(2\pi/L)\sin(\phi)$.

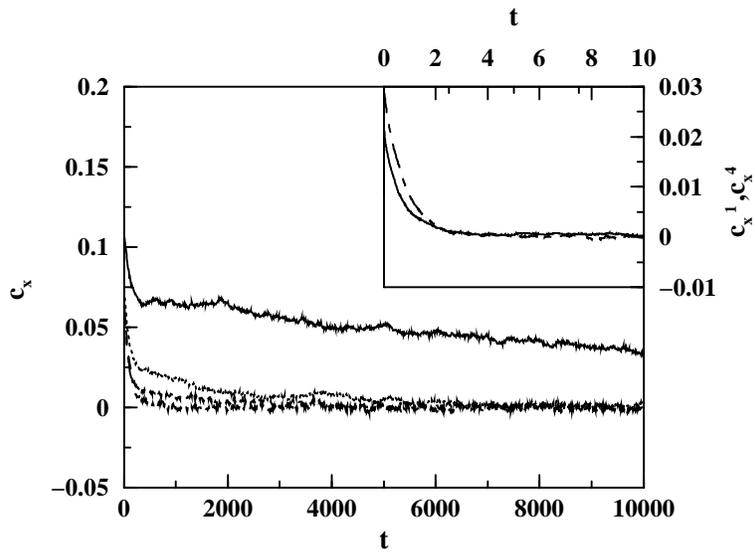
\noindent FIGURE 10: 
Velocity autocorrelation functions $c_x$ (in units of $U^2=25$) 
for a system with parameters
$L_x=20$, $L_y=20$, $N=50$, ($\bar{\nu}=0.426$). In order of
increasing values for short times are $e_n=0.8, 0.78, 0.77, 0.76$.
Inset: Velocity autocorrelation functions $c_x^i$ (same units) for a system
$L_x=80$, $L_y=20$, $N=200$, ($\bar{\nu}=0.426$) and $e=0.76$. The
system was divided in $5$ stripes of width $4$. Correlation functions
are shown only in the stripes at the walls. Solid line marks the function
at the wall of the cluster, and dot-dashed line at the opposite wall.
Time is measured in natural units ($r=1$, $U=5$).

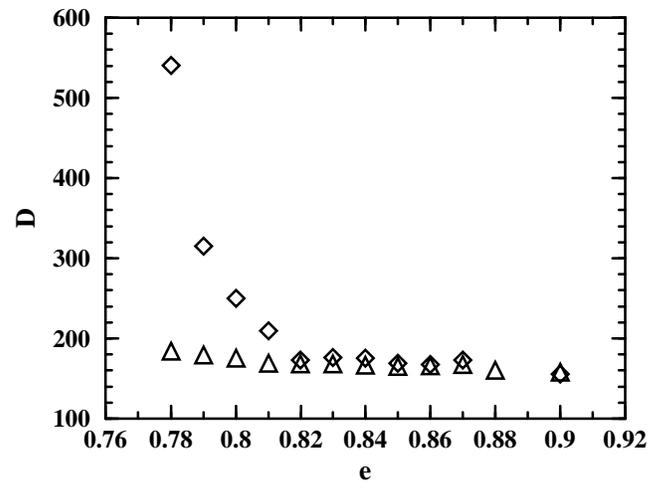
\noindent FIGURE 11: The self-diffusion coefficient as a function
of the restitution coefficient $e_n$ for $L_x = L_y = 20$, $N=50$, 
($\bar{\nu}=0.426$) measured in natural units. This runs were carried 
out moving both walls in
opposite direction with $U=\pm5$. Diamonds ($\Diamond$) marking values
determined from the mean square displacement, and triangles
($\triangle$) marking values determined from the velocity
autocorrelation function.

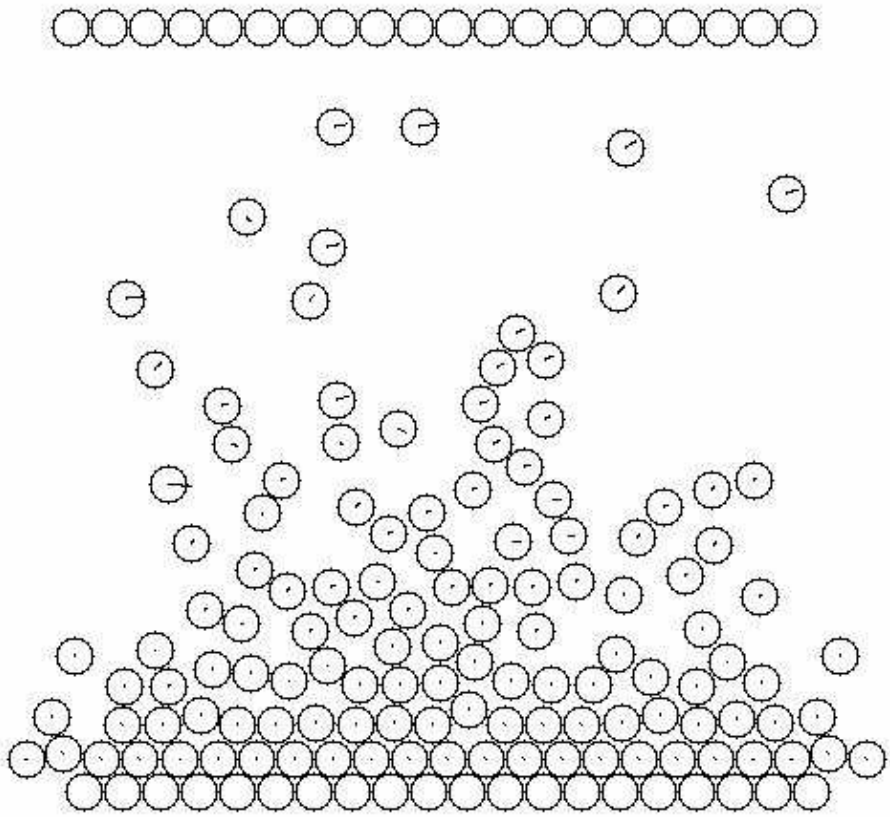
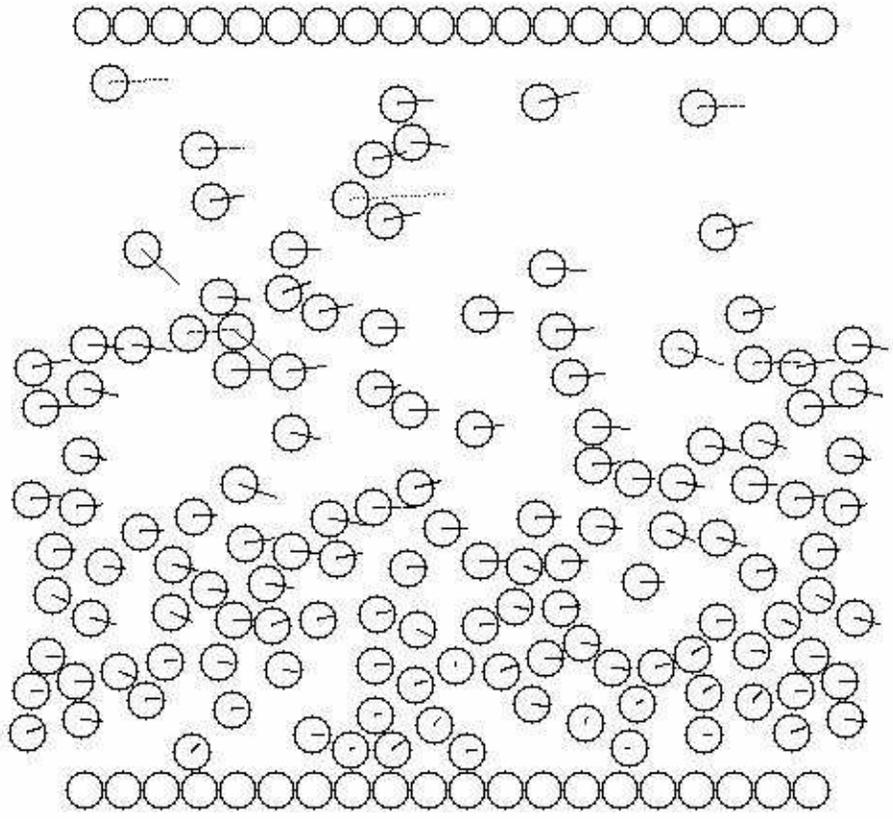
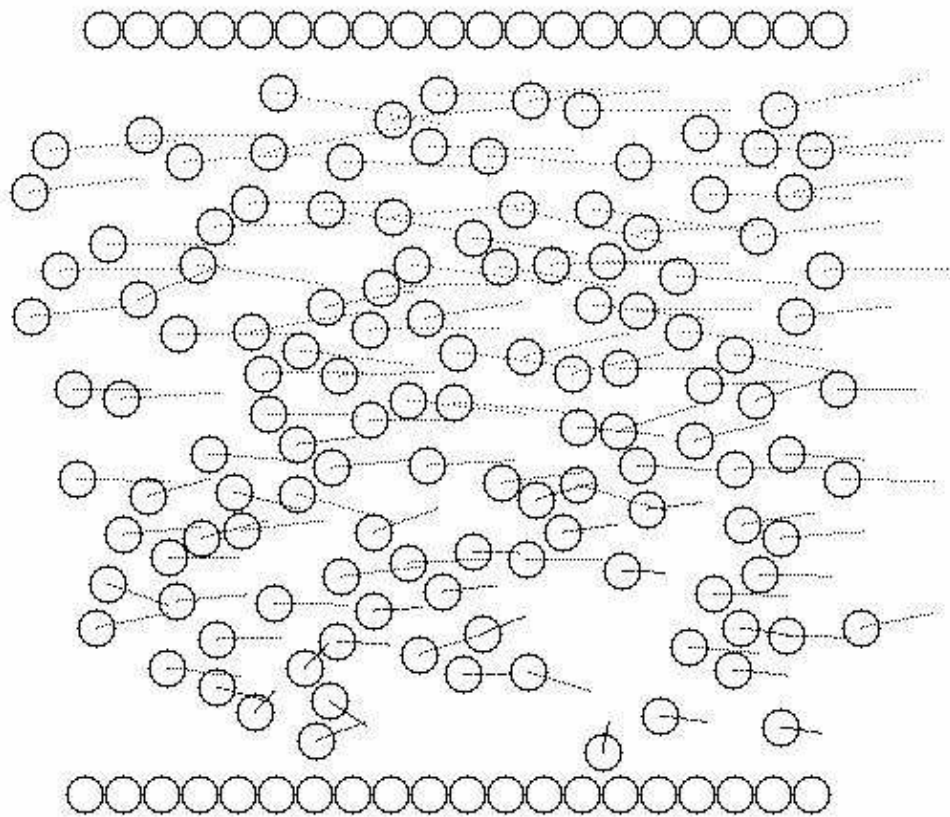
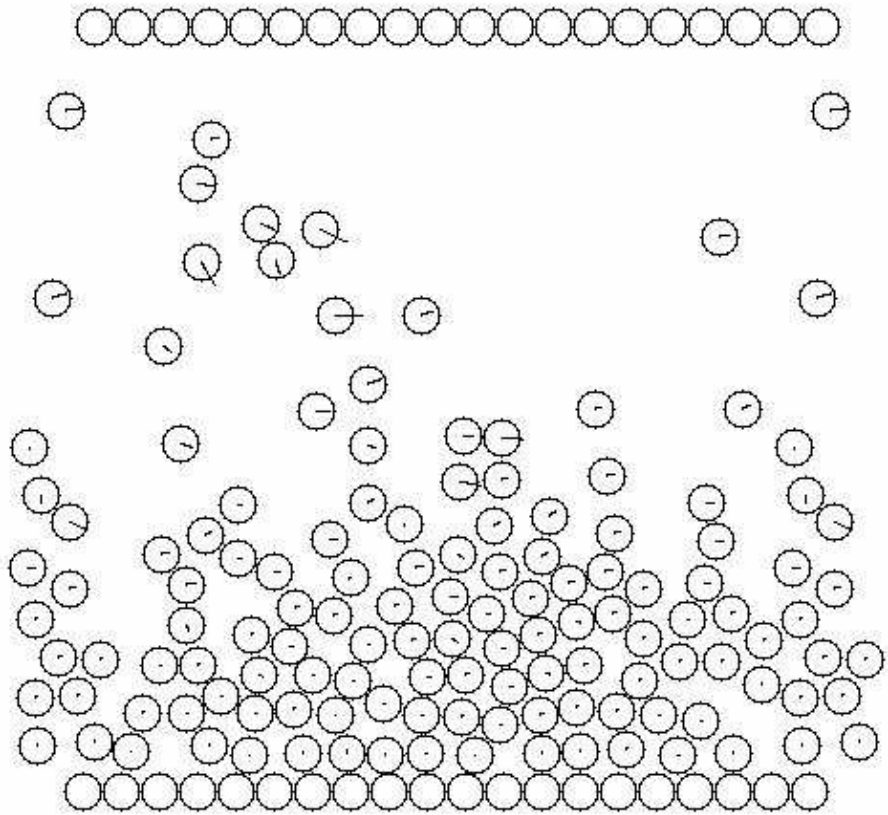
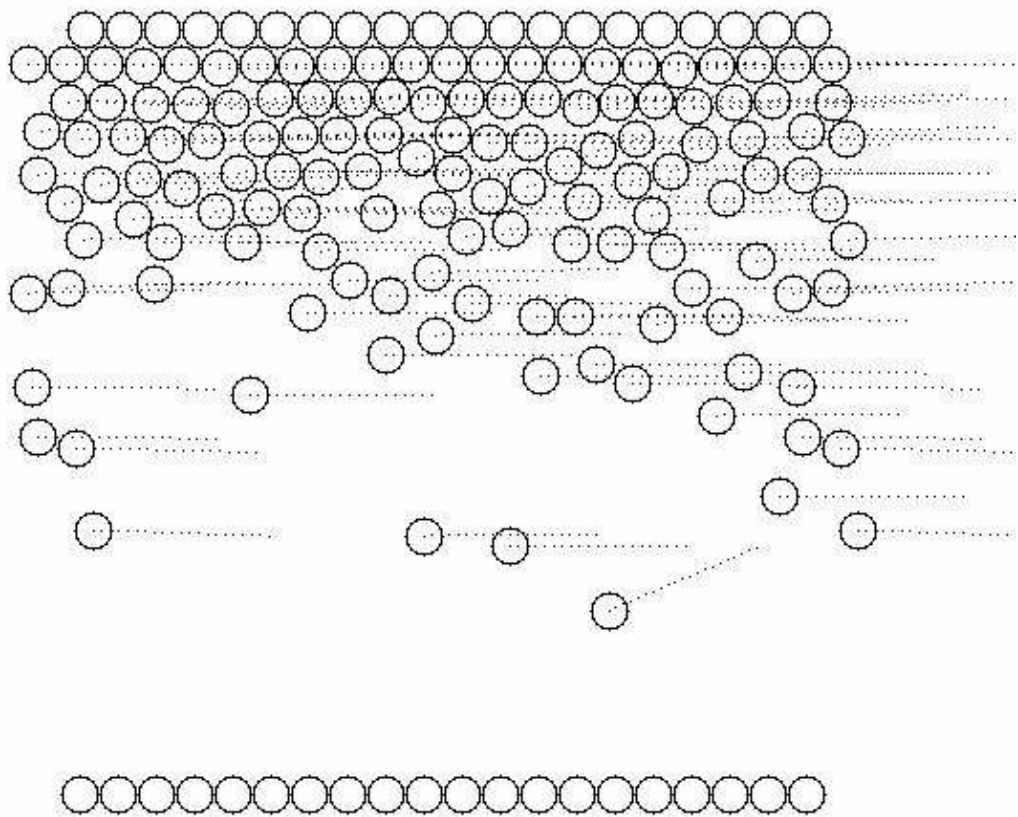
\end{document}